\documentclass[12pt]{elsarticle}

\makeatletter
\def\ps@pprintTitle{%
 \let\@oddhead\@empty
 \let\@evenhead\@empty
 \def\@oddfoot{\centerline{\thepage}}%
 \let\@evenfoot\@oddfoot}
\makeatother

\usepackage{amsmath}
\usepackage{amssymb}
\usepackage{graphicx}
\usepackage{amsfonts}
\usepackage{mathtools}
\usepackage{mathrsfs}

\usepackage{algorithm}
\usepackage{algorithmic}
\usepackage{accents}
\usepackage{multirow}

\usepackage{multirow}
\usepackage{epstopdf}
\usepackage{adjustbox}

\usepackage{tikz}
\usetikzlibrary{arrows,backgrounds,snakes}

\usepackage{exscale,relsize} 

\newcommand{\dist}{{\textstyle \mathsmaller{\mathit{\Delta}}}} 
\renewcommand{\vec}[1]{\mathbf{vec}(#1)}

\begin{document}

\begin{frontmatter}



\title{A Bag-of-Paths Node Criticality Measure}


\author{Bertrand~Lebichot\footnote{Corresponding author. Tel.: +32 10 47 83 77.\\
E-mail addresses:\\
bertrand.lebichot@uclouvain.be (B. Lebichot),\\
marco.saerens@uclouvain.be (M. Saerens).\\
URL:\\
http://www.isys.ucl.ac.be/staff/lebichot/ (B. Lebichot)\\
http://www.isys.ucl.ac.be/staff/marco/index.htm. (M. Saerens)}
        \& Marco~Saerens}

\address{Machine Learning Group - ICTEAM \& LSM, Universit\'e catholique de Louvain \\ Place des Doyens 1, B-1348 Louvain-la-Neuve, Belgium }

\begin{abstract}
This work compares several node (and network) criticality measures quantifying to which extend each node is critical with respect to the communication flow between nodes of the network, and introduces a new measure based on the Bag-of-Paths (BoP) framework. Network disconnection simulation experiments show that the new BoP measure outperforms all the other measures on a sample of Erd\H os-R\'{e}nyi and Albert-Barab\'asi graphs. Furthermore, a faster (still $O(n^{3}))$, approximate, BoP criticality relying on the Sherman-Morrison rank-one update of a matrix is introduced for tackling larger networks. This approximate measure shows similar performances as the original, exact, one.
\end{abstract}


\end{frontmatter}


\section{Introduction}
%
%
%
%

The analysis and the modeling of network data has become a popular research topic in the last decade and is now often referred to as \emph{link analysis} (in computer science) and \emph{network science} (in physics). Network data appear in virtually every field of science and is therefore studied in many different disciplines, such as social sciences, applied mathematics, physics, computer science, chemistry, biology, economics, etc.
Within this context, one important question that is often addressed is the following: Which node seems to be the most critical, or vital, in the network?
The present work introduces such a new \emph{node criticality measure}, also called \emph{vulnerability}, quantifying to which extend the deletion of each node hurts the connectivity within the network in a broad sense, e.g., in terms of communication, proximity, or movement. Criticality measures are often considered as a subset of centrality measures, which are frequently used as a proxy for quantifying criticality. Interested readers are invited to consult the recent comprehensive review \cite{Lu-2016}.

Indeed, a huge number of centrality measures have been defined in various fields, starting from social science (see, e.g., \cite{Freeman-1977,Newman-05,Estrada-2009,Wehmuthr-2011,Klein-2010,Tizghadam-2010} and \cite{Brandes-2005} for a survey). These quantities assign a score to each node of the graph $G$ which reflects the extent to which this node is ``central" by exploiting the structure of the graph $G$, or with respect to the communication flow between nodes. Centrality measures tend to answer the following questions ADD REF: What is the most representative, or central, node within a given graph (closeness centrality)? How critical is a given node with respect to the information flow in a network (criticality)? Which node is the most peripheral in a social network (eccentricity)? Which node is the most important intermediary in the network (betweenness centrality)? Centrality scores try to answer to these questions by proposing measures modeling and quantifying these different, somewhat vague, properties of the nodes.

Notice that, in general, these centrality measures are computed on undirected graphs, or, when dealing with a directed graph, by ignoring the direction of edges. They are therefore denoted as ``undirectional" \cite{Wasserman-1994}. Measures defined on directed graphs -- and therefore directional -- are often called \emph{importance} or \emph{prestige} measures. They capture to which extend a node is ``important", ``prominent", or ``prestigious" with respect to the whole directed graph by considering directed edges as representing some kind of endorsement. However, this kind of measure will not be discussed here.

This work introduces a new, efficient and effective, criticality measure: the bag-of-paths (BoP) criticality. The quantity relies on the bag-of-paths framework assigning a Gibbs-Boltzmann distribution on the set of paths in the network \cite{Francoisse-2012,MOI,Devooght-2014}. This framework already allowed to define new distance measures between nodes interpolating between two well-known distances, the shortest-path distance and the resistance distance (or commute-time distance) \cite{Francoisse-2012}. In this context, the BoP criticality of a node measures the impact of the node deletion on the total accessibility between nodes within the network. More specifically, it is defined as the Kullback-Leibler divergence between the bag-of-paths probabilities, quantifying relative accessibilities, computed before and after removal of a node of interest. The larger this decrease in accessibility, the higher the impact of the node deletion, and thus the higher its criticality.

The novelty of the approach introduced in this paper can be understood as follows. Most of the traditional criticality measures are essentially based on two different paradigms about the communication occurring in the network: optimal communication based on shortest paths and random communication based on a random walk on the graph. For instance, the Wiener index (described later in this paper) is based on shortest paths and the Kirchhoff index on a random walk. However, both the shortest path and the random walk have some drawbacks: shortest paths do not integrate the amount of connectivity between the two nodes whereas random walks loose the notion of proximity to the initial node when the graph becomes larger \cite{vonLuxburg-2010}. Contrary to traditional measures, our criticality measure integrates both proximity and amount of connectivity in the bag-of-paths framework. Nodes that are both close and highly connected are qualified as highly \emph{accessible}. Our introduced bag-of-paths measure aims to quantify the accessibility between the nodes. When the temperature of the model is low (close to zero), communication occurs through a random walk, while for large temperatures, short paths are promoted.

The introduced measure is compared experimentally to already developed criticality measures as well as a sample of popular centrality measures,
briefly reviewed in this paper. All those measures are compared through a Kendall's correlation analysis and a \emph{disconnection methodology} \cite{albert-2000error,holme-2002} in Section \ref{Exp}. This empirical analysis is performed on a large number, and two types, of randomly generated graphs (see Subsection \ref{Data}).

In summary, this work has the following main contributions,
\begin{itemize}
		\item A new criticality measure, showing good performance in the identification of the most critical nodes of a network, is introduced.
		\item All those methods are compared experimentally using two disconnection strategies on a large number of randomly generated graphs.
	\end{itemize}

Finally, the paper is organized as follows: First, the underlying background and various notations are discussed in Section \ref{BgNot}, then Section \ref{RelW} introduces ten centrality and criticality measures (some being quite well-known). The bag-of-paths model described in \cite{Francoisse-2012} is summarized and the new BoP criticality measure is derived in Section \ref{BoPC}. Finally, those measures are assessed and compared in Section \ref{Exp}.

\section{Background and Notation}
\label{BgNot}

This section aims to introduce the necessary background and notation used in this paper. 
Consider a weighted directed graph or network, $G = \{ \mathcal{V} ,\mathcal{E} \}$, strongly connected  with a set of $n$ nodes $\mathcal{V}$ (or vertices) and a set of edges $\mathcal{E}$ (or arcs, links). The $n\times n$ adjacency matrix of the graph, containing non-negative affinities between nodes, is denoted as $\mathbf{A}$, with elements $[\mathbf{A}]_{ij} = a_{ij} \ge 0$.

$\mathbf{A}^\text{T}$ will refer to the transpose of $\mathbf{A}$, $\mathbf{A}^{(-j)}$ is a $(n-1) \times (n-1)$ matrix obtained from $\mathbf{A}$ by removing its $j$th row and its $j$th column, $\mathbf{e}$ is a column vector full of ones and $\mathbf{e}_{j}$ is the $j$th column vector of the identity matrix $\mathbf{I}$. Except explicitly stated, all lower-case bold letters represent column vectors while upper-case bold letters are matrices.

Moreover, to each edge between node $i$ and $j$ is associated a non-negative number $c_{ij}\ge 0$. This number represents the immediate cost of transition from node $i$ to $j$. If there is no link between $i$ and $j$, the cost is assumed to take a large value, denoted by $c_{ij} = \infty$. The cost matrix $\mathbf{C}$ is an $n\times n$ matrix containing the $c_{ij}$ as elements. Costs are usually set independently of the adjacency matrix: they are quantifying the cost of a transition according to the problem at hand. For example, costs can be set in function of some properties, or features, of the nodes (or the edges) in order to bias the probability distribution of choosing a path to follow. In the case of a social network, we may, for instance, want to bias the paths in function of the education level of the persons, therefore favoring paths visiting highly educated persons. Now, if there is no reason to introduce a cost, we can simply set $c_{ij} = 1$ (paths are penalized by their length) or $c_{ij} = 1/a_{ij}$ (in this case, $a_{ij}$ is viewed as a conductance and $c_{ij}$ as a resistance) -- this last setting will be used in the experimental section.

We also introduce the Laplacian matrix $%
\mathbf{L}$ of the graph, defined in the usual manner and needed below,
\begin{equation}
\mathbf{L}=\mathbf{D}-\mathbf{A}  \label{Eq_laplacian01} 
\end{equation}where
$\mathbf{D} = \mathbf{Diag}(\mathbf{A}\mathbf{e})$ is
the diagonal (out)degree matrix of the graph $G$ containing the $a_{i \bullet}$ on its diagonal.
One interesting property of $\mathbf{L}$ is that its eigenvalues provide important information about the connectivity of the graph \cite{Chung-1997}.

One of the most interesting accessibility measure of the graph $G$, the so-called connectivity, is often defined as the minimum number of nodes that need to be removed to separate it into two disconnected sub-graphs \cite{Harary-1969,Seidman-1983}. Unfortunately, this quantity is hard to compute and cannot be easily exploited in practice for this reason. Beside this, it can be shown that the number of zero eigenvalues of $\mathbf{L}$ is equal to the number of disconnected subgraphs, or connected components, of $G$ \cite{Chung-1997}. Then, for a connected graph the smallest eigenvalue of $\mathbf{L}$ is called the algebraic connectivity or spectral gap and has been shown to be a good indicator of its overall ``connectedness" ($G$ is disconnected when its algebraic connectivity is equal to zero). Finally, the Moore-Penrose pseudoinverse of $\mathbf{L}$ is denoted as $\mathbf{L}^+$, and contains elements $l^{+}_{ij}$. Due to the properties of the Moore-Penrose pseudoinverse, its largest eigenvalue is the algebraic connectivity.

In addition, a natural random walk on $G$ is defined in the standard way. In node $i$, the random walker chooses the next edge to follow according to reference transition probabilities
\begin{equation}
	p_{ij}^{\mathrm{ref}}=\frac{a_{ij}}{{\sum_{j'=1}^{n}}a_{ij'}}
	\label{Pref}
\end{equation}
The $n\times n$ matrix $\mathbf{P}^{\mathrm{ref}}$, containing transition probabilities $p_{ij}^{\mathrm{ref}}$, is stochastic and is simply equal to $\mathbf{P}^{\mathrm{ref}} = \mathbf{D}^{-1}\mathbf{A}$. Note that this can lead to a division by zero if a node $i$ is isolated or a dangling node; we therefore assume that the graph is strongly connected.  $\mathbf{P}^{\mathrm{ref}}$ represents the probability of jumping from any node $i$ to node $j \in \mathcal{S}ucc(i)$, the set of successor nodes of $i$.  In other words, the random walker chooses to follow an edge with a likelihood proportional to the affinity (apart from the sum-to-one normalization), therefore favoring edges with a large associated affinity. 

A path $\wp$ (also called a walk) is a sequence of transitions to adjacent nodes on $G$ (loops are allowed), initiated from a starting node $s$, and stopping in an ending node $e$. The total cost of a path $\wp$, $\tilde{c}(\wp)$, is defined as the sum of the individual transition costs $c_{ij}$ along $\wp$.  

\section{Related Work}
\label{RelW}

In this paper, a large set of criticality measures will be compared experimentally, and briefly reviewed in this section (see also \cite{Fouss-2016}). It is convenient to categorize them into three classes: node betweenness centrality measures, global graph criticality measures, and node criticality measures.

\subsection{Node betweenness centralities}
\label{Bet}

As already mentioned, the concept of criticality is closely related to the concept of betweenness centrality; we therefore also investigate a few of the most well-known betweenness and centrality measures. The measure is defined on each node, identified by its index $j$.
	
\begin{itemize}
		\item The simple \emph{node degree}, or \emph{edge connection} (EC). This quantity is simply the number of nodes connected to a node $j$, weighted by edge weights in the case of a weighted graph. It is obtained by summing the entries on the $j$th row of the adjacency matrix $\mathbf{A}$. The idea is that if a node has a high degree, it is more likely to hurt or disconnect the graph when removed. It can be computed by
\begin{equation}
\text{EC}_{j}
= \mathbf{e}_{j}^\text{T} \mathbf{A} \mathbf{e}
\label{EC}
\end{equation}
		
%

\item The famous \emph{shortest path betweenness} (SPB), introduced by Freeman \cite{Freeman-1977}. It counts the proportion
of shortest paths connecting any two nodes $i$ and $k$, and passing through an intermediate node $j$ of interest (with $ i  \neq j \neq k \neq i$). The idea is that if a node contributes to a large number of shortest paths, it can be considered as an important intermediary between nodes when the information is spread ``optimally" along shortest paths. More precisely,
\begin{equation}
\text{SPB}_{j}
= \displaystyle \sum_{\substack{i=1 \\ i \ne j}}^n \displaystyle \sum_{\substack{k=1 \\ k \ne i,j}}^n  \frac{\eta(j \in \mathcal{P}^*_{ik})}{|\mathcal{P}^*_{ik}|}
\label{SPB}
\end{equation}
where $\mathcal{P}^*_{ik}$ is the set of all shortest paths from $i$ to $k$, $|\mathcal{P}^*_{ik}|$ is the total number of such shortest paths $\wp_{ik}^{*}$ and $\eta(j \in \mathcal{P}^*_{ik}) = \sum_{\wp_{ik}^{*} \in \mathcal{P}^*_{ik}} \delta(j \in \wp_{ik}^*)$ is the total number of such paths visiting node $j$. We used Brandes' algorithm \cite{Brandes-2001} to compute the SPB of each node of the graph.


\item The \emph{random walk betweenness} (RWB), introduced by Newman \cite{Newman-05} and closely related to Brandes' electrical centrality \cite{Brandes-2005b}. Newman introduced the current flow betweenness centrality, which measures the centrality of a node as the total sum of electrical current that flow through it, when considering all node pairs as source-destination pairs with a unit current flow. The current flow betweenness is also called the \emph{random walk betweenness centrality} because of the well-known connection between electric current flows and random walks~\cite{Snell-1984,Fouss-2016}.
The idea is thus the same as for the SPB, but taking into account a random walk-based diffusion of information instead of shortest paths. Notice that Brandes and Fleischer \cite{Brandes-2005b} proposed a more efficient algorithm computing the random walk betweenness for all nodes of a network. The properties and computation of the current flow betweenness have also been discussed by Bozzo and Franceschet \cite{Bozzo-2013}. Kivimaki et al. proposed a new betweenness measure interpolating between the shortest path betweenness and the random walk betweenness \cite{Kivimaki-2016}.
%

		\item \emph{Estrada's centrality} (EST). In \cite{Estrada-2009}, Estrada et al. defined a centrality measure called ``subgraph centrality" for a weighted undirected graph or subgraph. It summarizes simply as
\begin{equation}
\text{EST}_{j} = \mathbf{e}_{j}^\text{T} \left( \sum\limits_{k=0}^{\infty }\frac{\mathbf{A}^{k}}{k!} \right) \mathbf{e}_{j} = \mathbf{e}_{j}^\text{T} \, \mathbf{diag}(\mathrm{expm}(\mathbf{A}))
\label{EST}
\end{equation}
where $\mathrm{expm}(\mathbf{A})$ is the matrix exponential of $\mathbf{A}$ and $\mathbf{diag}(\mathbf{X})$ extract the main diagonal of $\mathbf{X}$.
It is well-known that element $a_{ij}^{(k)}=[\mathbf{A}^{k}]_{ij}\mathbf{\ }$ of matrix $\mathbf{A}^{k}$ ($\mathbf{A}$ to the power $k$) is the weighted number of paths between node $i$ and node $j$ with exactly $k$ steps. The subgraph centrality measure therefore integrates a contribution from all paths connecting node $j$ to himself, discounting paths according to their number of steps (it favors shorter paths in terms of length). The intuition is that a node should have a high centrality score if the closed paths (cycles) starting from it are short and are visiting many different nodes \cite{Estrada-2009}.


%
	
\end{itemize}	

\subsection{Node criticalities}
\label{N Crit}

We now introduce the node criticalities studied in this work. As for the betweenness, the criticality measure is defined on each node $j$.

\begin{itemize}

		\item \emph{Wehmuth's criticality} $K$ (WK) is introduced in \cite{Wehmuthr-2011}, 
		
\begin{equation}
\text{WK}_{j}
= \frac{\lambda_2^{(j)}}{\log_2(d_j)}
\label{WK}
\end{equation}where $\lambda_2^{(j)}$ is the algebraic connectivity of the $h$-neighbourhood of node $j$ (the subnetwork composed by all nodes within $h$ hops of node $j$) and $d_j$ is the degree of node $j$. Recall that the algebraic connectivity is the second smallest eigenvalue of the Laplacian matrix $\mathbf{L}$. The idea is to take advantage of the algebraic connectivity property; the higher the value of $\lambda_2^{(j)}$, the higher the connectivity/density of the subnetwork. Then, $\lambda_2^{(j)}$ is divided by the logarithm of the node degree as locally computed algebraic connectivities show a bias towards higher values on nodes with high degree. This bias causes $\lambda_2^{(j)}$ to be over-sensitive to the presence of hubs \cite{Wehmuthr-2011}.

	
		\item \emph{Klein's edge criticality} (KLE). Klein derived the analytical form of this node criticality measure for several global measures, including the Wiener index and the Kirchhoff index \cite{Klein-2010}. We will use the measure based on the Kirchhoff index here \cite{Klein-2010},
\begin{equation}
\text{KLE}_j
= \displaystyle \sum_{i=1}^n a_{ij} (\mathbf{e}_{i}-\mathbf{e}_{j})^\text{T} (\mathbf{L}^+)^2 (\mathbf{e}_{i}-\mathbf{e}_{j})
\label{KLE}
\end{equation}
The intuition behind the measure is the following. Klein's edge $(i,j)$ criticality is defined as the sensitivity of the global network criticality index (here the Kirchhoff index -- defined in the next subsection) with respect to the increase in the resistance of the edge $(i,j)$ \cite{Klein-2010}.
In other words, it quantifies the impact of an increase in this resistance on the global network. Edges having a high impact on the global network criticality hurt most the network and are considered as highly critical. Then, edge criticality is summed up over incident edges to provide a node criticality.

%

\end{itemize}

\subsection{Global network criticalities}
\label{G Crit}

The following \emph{global} criticality indexes are defined on the whole network $G$. They quantify the extend to which the network as a whole is efficient, that is, highly interconnected and cohesive, with high accessibility. For a communication network, this measure can be, e.g., the ``Wiener index" -- the sum of the shortest-path distances (which can be travel time, travel cost, etc.) between all pairs of nodes.
An effective network is characterized by a \emph{low value} of the Wiener index as, then, distances between nodes are small in average.

The impact of a node of interest on the global network accessibility measure -- the \emph{derived node criticality} -- is then quantified by evaluating the marginal loss in global accessibility when the node of interest is not operating, i.e., has simply been removed.
This measure therefore reports how critical the node is, relative to the entire graph. To evaluate the criticality of a particular node $j$ in a fixed graph $G$, the difference between the global criticality after deleting this node $j$, $\text{cr}(G \setminus j)$, and the initial global network criticality, $\text{cr}(G)$, is computed,
\begin{equation}
\text{cr}_{j} = \text{cr}(G \setminus j) - \text{cr}(G)
\label{Eq_global_graph_criticality}
\end{equation}

This node criticality will be computed on several well-known global criticality measures which are described now. We could also normalize the quantity when it corresponds to a sum over all pairs of nodes by something like $ \mathrm{cr}(G)/(n(n-1)) - \mathrm{cr}(G \setminus i)/((n-1)(n-2)) $. However, this would not change the ranking of the nodes as the first term is a constant.

\begin{itemize}	

		\item The \emph{Wiener index} (WIE) is defined as the sum of the shortest-path distances between all node pairs (see, e.g., \cite{Brandes-2005}),
\begin{equation}
\text{WIE}(G)
= \frac{1}{2} \displaystyle \sum_{i=1}^n \displaystyle \sum_{j=1}^n \dist_{ij}^{\mathrm{SP}}
\label{WIE}
\end{equation}where $\dist_{ij}^{\mathrm{SP}}$ is the shortest-path distance. The underlying idea is that if the sum of the distances between every node pairs is small, the network is more likely to be well-connected.

		\item The \emph{Kirchhoff index} (KIR) is similar to the Wiener index but uses the resistance distance (the effective resistance, proportional to the commute-time distance based on a random walk on the graph) \cite{Klein-1993}, instead of the shortest path distance, and has been recently used by Tizghadam and al. in network theory for quantifying the robustness of a communication network \cite{Tizghadam-2010}. It can be easily computed by
\begin{equation}
\text{KIR}(G)
= \frac{1}{2} \displaystyle \sum_{i=1}^n \displaystyle \sum_{j=1}^n \dist_{ij}^{\mathrm{ER}}
\label{KIR}
\end{equation}
where $\dist_{ij}^{\mathrm{ER}}$ is now the effective resistance between $i$ and $j$.
The idea is thus the same as for WIE, but with a different concept of distance.

		\item The \emph{Kemeny index} (KEM) represents the expected number of steps needed by a random walker for reaching an arbitrary node from some arbitrary starting node \cite{Kemeny-1976}, when the starting and ending nodes are selected according to the equilibrium distribution of the Markov chain. Indeed, for an irreducible, aperiodic, Markov chain, it is known 
(see, e.g., \cite{Norris-1997}) that the stationary distribution exists and is independent of the initial state $i$. More precisely, the Kemeny index is
\begin{equation}
\text{KEM}(G)
= \displaystyle \sum_{i=1}^n \pi_i \displaystyle \sum_{j=1}^n  \pi_j m_{ij} = \displaystyle \sum_{j=1}^n \pi_j m_{ij}
\label{KEM}
\end{equation}where $m_{ij}$ is the average first-passage time between node $i$ and node $j$ and  $\boldsymbol{\pi}$ is the stationary distribution. Equation (\ref{KEM}) holds because it can be shown that the quantity $\sum_{j=1}^{n} \pi_{j} m_{ij}$ is independent of the starting node $i$ \cite{Doyle-2009}. This index measures the relative accessibility of all pairs of nodes, putting more weight on the long-term frequently visited nodes according to the stationary distribution.

	
	\item The \emph{Shield value} (SHV) has recently been introduced \cite{Tong-2010}:

\begin{equation}
\text{SHV}(G)
= \lambda_{1}
\label{SHV}
\end{equation}
where $\lambda_{1}$ is the dominant eigenvalue of the adjacency matrix $\mathbf{A}$.
It is closely related to the loop capacity and the path capacity of the graph, that is, the number of loops and paths of finite length. The higher $\lambda_{1}$, the more loops and long path in the graph. As for Estrada's centrality, the underlying idea is that if a graph has many such loops and paths then it is more likely to be well connected. The more the deletion of a node lowers this value, the less the graph becomes connected, and therefore the larger its criticality value.

\end{itemize}

		
\section{The Proposed BoP Criticality}
\label{BoPC}		
		
We now derive a new node criticality measure called the bag-of-paths criticality (BPC). It is based on computing the effect of a node removal in a bag-of-paths model (BoP). This framework was recently introduced in \cite{Francoisse-2012} (see also \cite{Mantrach-2009} for a related work) for computing distances on graphs, and used for semi-supervised classification tasks in \cite{Francoisse-2012,MOI}. In order to make the paper as self-contained as possible, we briefly review this framework first in this section. Finally, an illustrative example is shown in Subsection \ref{Ill}.

\subsection{The bag-of-paths model}

The BoP framework is based on the probability of drawing a path $i \leadsto j$ starting at a node $i$ and ending in a node $j$ from a virtual bag containing all possible paths in the network \cite{Francoisse-2012}. Let us define $\mathcal{P}_{ij}$ as the set of all paths connecting node $i$ to node $j$, including loops. We further define the set of all paths through the network as $\mathcal{P}=\bigcup_{i,j=1}^{n}\mathcal{P}_{ij}$.

The potentially infinite set of paths in the graph is enumerated and a probability distribution is assigned to the set of individual paths $\mathcal{P}$, considered independently. This probability distribution on the set $\mathcal{P}$ represents the probability of drawing a path $\wp\in\mathcal{P}$ from the bag, and is defined as the probability distribution $\text{P}(\cdot)$ minimizing the total expected cost along path $\wp$,
$\mathbb{E}\left[ \tilde{c}(\wp)\right]$, among all the distributions having a fixed relative entropy $J_0$ with respect to a reference distribution, for instance the natural random walk on the graph (defined by Equation (\ref{Pref})). The quantity $\tilde{c}(\wp)$ is the cumulated cost along path $\wp$.

This choice naturally defines a probability distribution on the set of paths such that ``long" (high cost) paths
occur with a low probability while ``short" (low cost) paths occur with a high probability.
In other words, we are seeking path probabilities, $\textnormal{P}(\wp), \wp\in\mathcal{P}$,
minimizing the total expected cost subject to a constant relative
entropy constraint,
\begin{equation}
\vline\,\begin{array}{llll}
\underaccent{\{\textnormal{P}(\wp)\}}{\mathrm{minimize}} & {\displaystyle \sum_{\wp\in\mathcal{P}}}\text{P}(\wp) \, \tilde{c}(\wp)\\[0.5cm]
\mathrm{subject\,to} & \sum_{\wp\in\mathcal{P}}\textnormal{P}(\wp)\ln(\textnormal{P}(\wp)/\tilde{\pi}^{\mathrm{ref}}(\wp))=J_{0} \\ & \sum_{\wp\in\mathcal{P}}\textnormal{P}(\wp)=1
\end{array}
\label{BoPmini}
\end{equation}
where $\tilde{\pi}^{\mathrm{ref}}$ represents the probability of following 
the path $\wp$ when walking according to the natural random walk reference distribution.
In other words, $\tilde{\pi}^{\mathrm{ref}}$ is the product of the transition probabilities $p_{ij}^{\mathrm{ref}}$ along the path $\wp$ -- the likelihood of the path.
Here, $J_{0} > 0$ is provided a priori by the user, according to the desired
degree of randomness, or exploration, he is willing to concede.

As well-known (see, e.g., ADD REF,\cite{ Francoisse-2012,Mantrach-2009,Saerens-2008} for details), this problem is similar to a standard maximum entropy one and can be solved by introducing the following Lagrange
function
\begin{equation}
\mathscr{L} = \sum_{\wp\in\mathcal{P}}\text{P}(\wp)\tilde{c}(\wp) + \lambda\left[\sum_{\wp\in\mathcal{P}}\text{P}(\wp)\ln \left( \frac{\text{P}(\wp)}{\tilde{\pi}^{\mathrm{ref}}(\wp)} \right) - J_{0}\right]\nonumber + \mu\left[\sum_{\wp\in\mathcal{P}}\text{P}(\wp)-1\right]
\end{equation}
and optimizing over the set of path probabilities $\{ \text{P}(\wp) \}_{\wp\in\mathcal{P}}$ (partial derivatives set to zero). The Lagrange parameters are then deduced after imposing the constraints.

The result of the minimization of (\ref{BoPmini}) is a Gibbs-Boltzmann probability distribution:
\begin{equation}
\text{P}(\wp) = \frac{\tilde{\pi}^{\mathrm{ref}}(\wp)\exp\left[-\theta \tilde{c}(\wp)\right]}{\displaystyle \sum\limits_{\wp' \in \mathcal{P}}\tilde{\pi}^{\mathrm{ref}}(\wp') \exp[-\theta \tilde{c}(\wp')] }
\label{Boltzmann}
\end{equation}
where $\theta = 1/T$ plays the role of an inverse temperature and $\exp$ is the elementwise exponential.
As expected, short paths $\wp$ (having a low $\tilde{c}(\wp)$) are favoured
in that they have a larger
probability of being chosen. Moreover, from Equation (\ref{Boltzmann}),
we clearly observe that when $\theta \rightarrow 0^{+}$, paths probabilities reduce
to the probabilities generated by the natural
random walk on the graph. In this case, $J_0 \rightarrow 0$ and paths are chosen according to their likelihood in a natural random walk.
On the other hand, when $\theta$ is large, the probability distribution defined by
Equation (\ref{Boltzmann}) is biased towards short paths
(shortest ones are more likely).
Notice that, in the sequel, it will be assumed that the user provides the
value of the parameter $\theta$ instead of $J_0$, with $\theta > 0$.

The \underline{bag-of-paths probability} \cite{Francoisse-2012}, $\text{P}(s = i , e = j)$, is an important quantity defined on the set of (starting, ending) nodes of the paths. It corresponds to the probability of drawing a path starting in node $i$ and ending in node $j$ from the virtual bag-of-paths:
\begin{equation}
			\text{P}(s = i , e = j) = 
			\frac		{\displaystyle \sum\limits_{\wp \in \mathcal{P}_{ij}}\tilde{\pi}^{\mathrm{ref}}(\wp) \exp[-\theta \tilde{c}(\wp)] }  
							{\displaystyle \sum\limits_{\wp' \in \mathcal{P}}\tilde{\pi}^{\mathrm{ref}}(\wp') \exp[-\theta \tilde{c}(\wp')] }
			\label{BagOfP}
\end{equation}
where $\mathcal{P}_{ij}$ is the set of paths connecting the starting node
$i$ to the ending node $j$.

In \cite{Francoisse-2012}, it is shown that this probability can be computed in matrix form by
\begin{equation}
\text{P}(s=i,e=j) = \frac{z_{ij}}{\displaystyle \sum_{i',j'=1}^{n} z_{i'j'}} , \text{ with } \mathbf{Z}=(\mathbf{I}-\mathbf{W}\mathbf{)}^{-1}
\label{Eq_bag_of_paths_probabilities}
\end{equation}where $z_{ij}$ is the element $i$, $j$ of matrix $\mathbf{Z}$, called the fundamental matrix and
\begin{equation}
	\mathbf{W} = \mathbf{P}^{\mathrm{ref}} \circ \exp[-\theta  \mathbf{C}]
	\label{W}
\end{equation}
with $\circ$ being the elementwise (Hadamard) product.

Notice that $\text{P}(s = i,e = j)$ is not symmetric. These probabilities quantify the \emph{relative accessibility} between the nodes and it was shown that minus their logarithm, $- \log \text{P}(s = i,e = j)$, defines a useful distance measure between nodes \cite{Francoisse-2012}. By construction this probability is high when the two nodes $i$ and $j$ are \emph{highly connected} (there are many terms in the numerator of Equation (\ref{BagOfP})) by \emph{low-cost} paths (each term of the numerator is large). In other words, it accurately captures the intuitive notion of relative accessibility. These BoP probabilities will serve as a basis for defining the BoP criticality.

Note that the BoP probabilities can also be used to define some betweenness measures \cite{Kivimaki-2016} which are related to well-known centrality/betweenness measures in some sense: if $\theta \rightarrow \infty$ the betweenness tends to be highly correlated with Freeman's betweenness \cite{Freeman-1977} (only shortest paths are considered), while if $\theta \rightarrow 0^+$, the betweenness tends to be highly correlated with Newman's betweenness \cite{Newman-05} (based on a natural random walk).

\subsection{The BoP criticality: basic, standard, case (BPC)}

We will now derive a closed-form formula for computing these probabilities when an intermediate node $j$ is deleted from the graph. Then, our BoP criticality measure for node $j$ will be the relative entropy (or Kullback-Leibler divergence) between the bag-of-paths probabilities -- the relative accessibility -- before and after removing node $j$ from $G$. It therefore quantifies to which extend the relative accessibility is affected by the deletion of node $j$.

The intuition is the following. The bag-of-paths criticality quantifies the global impact of a node deletion on the total relative accessibility of the nodes in the network
\begin{itemize}
\item by computing this accessibility \emph{before} and \emph{after node deletion},
\item and then by computing their difference by means of the Kullback-Leibler divergence.
\item This difference computes the \emph{loss in accessibility} when deleting each node in turn.
\end{itemize}
Thus, a critical node is a node whose deletion greatly affects the relative accessibility between the nodes. This criticality measure will be referred as BPC.

First, let us introduce some new notation. In Equation (\ref{Eq_bag_of_paths_probabilities}), $z_{ik}$ will be denoted as $z_{ik}(\mathbf{A})$ and $\mathbf{Z}$ as $\mathbf{Z}(\mathbf{A})$ since they are based on matrix $\mathbf{A}$. Then, $\mathbf{Z}^{(-j)}(\mathbf{A})$ is $\mathbf{Z}$ based on $\mathbf{A}$ (the original graph), but where the $j$th column and the $j$th row of $\mathbf{Z}$ \underline{have been removed}, and $z_{ik}^{(-j)}(\mathbf{A})$ with $i\ne j$ and $k\ne j$ is its $i$, $k$ element.

We further define $\text{P}_{ik}^{(-j)}(\mathbf{A}) = \text{P}^{(-j)}(s=i,e=k)$ based on the elements of $\mathbf{Z}^{(-j)}(\mathbf{A})$,
\begin{equation}
\text{P}_{ik}^{(-j)}(\mathbf{A}) =
  \frac{z_{ik}^{(-j)}(\mathbf{A})}{\displaystyle \sum_{\substack{
            i',k' = 1 \\
            i',k' \ne j}}^{n} z_{i'k'}^{(-j)}(\mathbf{A})}, \text{ with } i,k \ne j
\end{equation}
which corresponds to the BoP probabilities (see Equation (\ref{Eq_bag_of_paths_probabilities})) based on the whole original graph ($\mathbf{A}$), but where the support of the probability distribution is reduced to the set of nodes different from $j$ -- we do not consider node $j$ as potential source or destination node.

It is important not to confuse $\mathbf{Z}^{(-j)}(\mathbf{A})$ with $\mathbf{Z}(\mathbf{A}^{(-j)})$, which is $\mathbf{Z}$ based this time on $\mathbf{A}^{(-j)}$, where the $j$th column and the $j$th row of $\mathbf{A}$ have been removed (the graph with node $j$ deleted). Thus the $z_{ik}(\mathbf{A}^{(-j)})$ with $i\ne j$ and $k\ne j$ are the elements of $\mathbf{Z}(\mathbf{A}^{(-j)})$. Notice that $\mathbf{Z}^{(-j)}(\mathbf{A})$ and $\mathbf{Z}(\mathbf{A}^{(-j)})$ have the same size; both are $(n-1)\times(n-1)$ square matrices (node $j$ has been removed in both cases).

Furthermore, to compute $\mathbf{Z}(\mathbf{A}^{(-j)})$, let us now calculate the impact of the deletion of node $j$ on the fundamental matrix $\mathbf{Z}$, based on $\mathbf{A}^{(-j)}$.
In order to investigate the deletion of node $j$, the $j$th row of matrix $\mathbf{W}$ will be set to zero -- we cannot escape from $j$ any more (this is similar to set the $j$th row of matrix $\mathbf{A}$ to zero). This aims at transforming node $j$ into a killing, absorbing, node. The result is that all paths from $i$ to $k$ passing through node $j$ (with $i,k \ne j$) are eliminated from the set of paths $\mathcal{P}_{ik}$. We therefore define $\text{P}(s=i,e=k|\{ s,e \} \ne j)$ based on $\mathbf{A}^{(-j)}$ as
\begin{equation}
\text{P}_{ik}(\mathbf{A}^{(-j)}) =
  \dfrac{z_{ik}(\mathbf{A}^{(-j)})}{ \displaystyle \sum_{\substack{
            i',k' = 1 \\
            i',k' \ne j}}^{n} z_{i'k'}(\mathbf{A}^{(-j)})}, \text{ with } i,k \ne j
  \label{Eq_BoP_probabilities_without_node_j}
\end{equation}

Finally, the \textbf{bag-of-paths criticality} (BPC) is the Kullback-Leibler divergence between the bag-of-paths probabilities, quantifying relative accessibilities, before and after node removal,

\begin{equation}
\mathrm{cr}_j 
= {\displaystyle \sum_{\substack{
            i,k = 1 \\
            i,k \ne j}}^{n} \text{P}_{ik}^{(-j)}(\mathbf{A}) \, \log \left( \frac{\text{P}_{ik}^{(-j)}(\mathbf{A})}{\text{P}_{ik}(\mathbf{A}^{(-j)})} \right)
}
\label{Eq_bag_hitting_paths_criticality01}
\end{equation}
Note that computing the bag-of-paths criticality for all the $n$ nodes has a time complexity of about $O(n^3 + n (n-1)^3)$. The first term corresponds to the evaluation of $\text{P}^{(-j)}(\mathbf{A})$ (which requires a matrix inversion) and the second term to $n$ evaluations of $\text{P}(\mathbf{A}^{(-j)})$ (inversion of $n$ matrices, after deleting node $j$). This leads to an overall $O(n^4)$ time complexity.

\subsection{The BoP criticality: faster approximation (BPCf)}

In this subsection, we will modify the bag-of-paths criticality to obtain a $O(n^3 )$ time complexity instead of $O(n^4)$. It relies on the efficient approximation of the entries of
$\mathbf{Z}^{(-j)}$ in terms of the fundamental matrix $\mathbf{Z}=(\mathbf{I}-\mathbf{W})^{-1}$.
This version will be referred as BPCf.
 
Let us first define 
\begin{itemize}
\item $\mathbf{z}_{j}^{\mathrm{c}} = \mathbf{col}_j(\mathbf{Z}) = \mathbf{Z} \mathbf{e}_j$ and $\mathbf{z}_{j}^{\mathrm{r}} = \mathbf{row}_j(\mathbf{Z}) = \mathbf{e}_j^{\text{T}} \mathbf{Z} $ 
\item $\mathbf{w}_{j}^{\mathrm{c}} = \mathbf{col}_j(\mathbf{W}) = \mathbf{W} \mathbf{e}_j$ and $\mathbf{w}_{j}^{\mathrm{r}} = \mathbf{row}_j(\mathbf{W}) = \mathbf{e}_j^{\text{T}} \mathbf{W} $ 
\end{itemize}
where $\mathbf{col}_j$ and $\mathbf{row}_j$ are respectively the $j$th column (a column vector) and the  $j$th row (a row vector) of the matrix. 

Now, turning node $j$ into a killing, absorbing, node (no outgoing link from this node) can be achieved by defining a new matrix $\mathbf{W}^{(-j)} = \mathbf{W} - \mathbf{e}_j \mathbf{w}^{\mathrm{r}}_j$ since $\mathbf{W}$ is a Hadamard product between $\mathbf{P}^{\mathrm{ref}}$ and $\mathbf{C}$. Doing so, row $j$ is set to zero, meaning that node $j$ cannot be an intermediate node anymore. This corresponds to a rank-one matrix update.
We will now show that we obtain an extremely simple formula for the update of the fundamental matrix:
\begin{equation}
\mathbf{Z}(\mathbf{W}^{(-j)}) = ( \mathbf{I} - \mathbf{W}^{(-j)} )^{-1}
= \mathbf{Z} - \frac{\mathbf{z}_{j}^{\mathrm{c}} \mathbf{z}_{j}^{\mathrm{r}}}{z_{jj}}
\label{Eq_Sherman_criticality01}
\end{equation}
where only the entries $i,k \ne j$ of $\mathbf{Z}(\mathbf{W}^{(-j)})$ are meaningful.
Recall that $\mathbf{z}_{j}^{\mathrm{c}}$ is a column vector while $\mathbf{z}_{j}^{\mathrm{r}}$ is a row vector.

Indeed, this results from a simple application of the Sherman-Morrison formula (see, e.g., \cite{Golub-1996,Meyer-2000,Seber-2008}) for the inverse of a rank-one update of a matrix: if $\mathbf{c}$ and $\mathbf{d}$ are column vectors,
\begin{equation}
(\mathbf{B}+\mathbf{cd}^{\text{T}})^{-1}=\mathbf{B}^{-1}-\frac{\mathbf{B}^{-1}\mathbf{cd}^{\text{T}}\mathbf{B}^{-1}}{1+\mathbf{d}^{\text{T}}\mathbf{B}^{-1}\mathbf{c}} \label{Eq_Sherman_formula01}
\end{equation}
Now, from $\mathbf{W}^{(-j)} = \mathbf{W} - \mathbf{e}_j \mathbf{w}^{\mathrm{r}}_j$,
we have $(\mathbf{I} - \mathbf{W}^{(-j)}) = (\mathbf{I}-\mathbf{W})+\mathbf{e}_{j} \mathbf{w}_{j}^{\mathrm{r}}$.
By setting $\mathbf{B}^{-1} = \mathbf{Z}$, $\mathbf{B}=(\mathbf{I}-\mathbf{W})$, $\mathbf{c}=\mathbf{e}_{j}$
and $\mathbf{d}=(\mathbf{w}^{\mathrm{r}}_{j})^{\text{T}}$ in Equation (\ref{Eq_Sherman_formula01}),
we obtain
\begin{equation}
\mathbf{Z}(\mathbf{W}^{(-j)})
= ( \mathbf{I}-\mathbf{W}^{(-j)} )^{-1}
= \mathbf{Z}-\frac{\mathbf{Ze}_{j} \mathbf{w}_{j}^{\mathrm{r}}\mathbf{Z}}{{1 + \mathbf{w}_{j}^{\mathrm{r}}}\mathbf{Ze}_{j}}
\label{Eq_Sherman_applied01}
\end{equation}

Let us first compute the term $\mathbf{w}_{j}^{\mathrm{r}}\mathbf{Z}$ appearing both in the numerator and the denominator of the previous Equation (\ref{Eq_Sherman_applied01}). Since $\mathbf{Z} = (\mathbf{I} - \mathbf{W})^{-1}$, $(\mathbf{I} - \mathbf{W}) \mathbf{Z} = \mathbf{I}$, and thus
\begin{align}
\mathbf{w}_{j}^{\mathrm{r}} \mathbf{Z} &= ((\mathbf{w}_{j}^{\mathrm{r}})^{\text{T}} - \mathbf{e}_j + \mathbf{e}_j)^{\text{T}} \mathbf{Z} \nonumber \\
 &= -(\mathbf{e}_j - (\mathbf{w}_{j}^{\mathrm{r}})^{\text{T}})^{\text{T}} \mathbf{Z} + \mathbf{e}_j^{\text{T}} \mathbf{Z} \nonumber \\
 &= -\mathbf{e}_j^{\text{T}} + \mathbf{z}_{j}^{\mathrm{r}} = \mathbf{z}_{j}^{\mathrm{r}} - \mathbf{e}_j^{\text{T}}
 \label{Eq_Sherman_result01}
\end{align}

From Equation (\ref{Eq_Sherman_result01}), the denominator of the second term in the right-hand side of Equation (\ref{Eq_Sherman_applied01}) is
\begin{equation}
1 + \mathbf{w}_{j}^{\mathrm{r}}\mathbf{Ze}_{j} = 1 + \left( \mathbf{z}_{j}^{\mathrm{r}} - \mathbf{e}_j^{\text{T}} \right) \mathbf{e}_{j} = \mathbf{z}_{j}^{\mathrm{r}} \mathbf{e}_{j} = z_{jj}
 \label{Eq_Sherman_result02}
\end{equation}

Moreover, also from (\ref{Eq_Sherman_result01}), the numerator of the second term in the right-hand side of Equation (\ref{Eq_Sherman_applied01}) is
\begin{equation}
\mathbf{Ze}_{j} \mathbf{w}_{j}^{\mathrm{r}}\mathbf{Z}
= \mathbf{z}_{j}^{\mathrm{c}} \left( \mathbf{z}_{j}^{\mathrm{r}} - \mathbf{e}_j^{\text{T}} \right)
 \label{Eq_Sherman_result03}
\end{equation}

We substitute the results (\ref{Eq_Sherman_result02}) and (\ref{Eq_Sherman_result03}) in the denominator and the numerator of Equation (\ref{Eq_Sherman_applied01}), providing
\begin{equation}
\mathbf{Z}(\mathbf{W}^{(-j)}) 
= \mathbf{Z} - \frac{\mathbf{z}_{j}^{\mathrm{c}} \left( \mathbf{z}_{j}^{\mathrm{r}} - \mathbf{e}_j^{\text{T}} \right)}{z_{jj}}
\label{Eq_Sherman_result04}
\end{equation}

However, row and column $j$ should neither be taken into account, nor used, and can therefore be put to zero.
Indeed, since the last term of the numerator in Equation (\ref{Eq_Sherman_result04}), $\mathbf{z}_{j}^{\mathrm{c}} \mathbf{e}_{j}^{\text{T}}$, only updates the $j$th column, it can safely be ignored (this column $j$ is useless and will never be used), resulting in redefining the quantity as
\begin{equation}
\mathbf{Z}(\mathbf{W}^{(-j)}) = ( \mathbf{I} - \mathbf{W}^{(-j)} )^{-1}
= \mathbf{Z} - \frac{\mathbf{z}_{j}^{\mathrm{c}} \mathbf{z}_{j}^{\mathrm{r}}}{z_{jj}}
\nonumber
\end{equation}
and now the $j$th row as well as the $j$th column of $\mathbf{Z}(\mathbf{W}^{(-j)})$ are equal to zero. Indeed, elementwise, this last equation reads $z_{ik}(\mathbf{W}^{(-j)}) = z_{ik} - z_{ij}z_{jk}/z_{jj}$, which is equal to zero both when $i=j$ and $k=j$. We therefore obtain exactly Equation (\ref{Eq_Sherman_criticality01}).
Thus, the fundamental matrix $\mathbf{Z}$ needs to be inverted \emph{only once} and the elements $z_{ik}(\mathbf{A}^{(-j)})$ in Equation (\ref{Eq_BoP_probabilities_without_node_j}) are approximated by $z_{ik}(\mathbf{W}^{(-j)})$.

The resulting matrix has a $j$th row as well as a $j$th column equal to zero and it can be shown that each element $z_{ik}(\mathbf{W}^{(-j)})$ of $\mathbf{Z}(\mathbf{W}^{(-j)})$ corresponds to
\begin{equation}
z_{ik}(\mathbf{W}^{(-j)}) = {\displaystyle \sum_{\wp \in \mathcal{P}_{ik}^{(-j)}}} \tilde{\pi}^{\mathrm{ref}}(\wp) \exp\left[-\theta c(\wp) \right]
\label{Eq_fundamental_matrix_criticality01}
\end{equation}
where $\mathcal{P}_{ik}^{(-j)}$ is the set of paths avoiding node $j$.

It should be noted that this procedure only computes an \emph{approximation} of the BoP probabilities $\text{P}_{ik}(\mathbf{A}^{(-j)})$ (defined in Equation (\ref{Eq_BoP_probabilities_without_node_j})) when removing an intermediate node $j$. Indeed, for computing the exact probabilities on the graph $G \setminus j$, the natural random walk transition probabilities (the reference probability matrix $\mathbf{P}^{\mathrm{ref}}$) should also be updated, as the edges entering node $j$ cannot be followed any more. In our procedure, these reference probabilities are not updated, causing some (usually small) disturbance in comparison with explicitly deleting the node $j$ and recomputing the probabilities (including transition probabilities) from this new graph $G \setminus j$.
Relative performances of the exact BoP criticality and the approximated criticality will be investigated in the experiments.

Note that the expression could be adapted to exactly reflect node deletion, but the update formula becomes much more complex and we did not observe any significant difference between the two approaches in our experiments (see the next section).

One way to render the procedure exact would be to instead minimize expected cost subject to a fixed \emph{entropy} constraint, instead of the \emph{Kullback-Leibler} divergence in Equation (\ref{BoPmini}). This results in redefining the $\mathbf{W}$ matrix as
\begin{equation}
\mathbf{W} = \exp[-\theta  \mathbf{C}]
\end{equation}
instead of (\ref{W}). This solves the problem of the $\mathbf{P}^{\mathrm{ref}}$ update since this transition matrix does not appear any more. However, experiments showed that this choice performs slightly worse (therefore not reported in the paper) than the approximate update introduced in this section.

The algorithm is detailed in Algorithm \ref{Alg_bag_of_paths_criticality01}, where the probabilities $\text{P}_{ik}^{(-j)}(\mathbf{A})$ and $\text{P}_{ik}(\mathbf{A}^{(-j)})$ (approximated by $\text{P}_{ik}(\mathbf{W}^{(-j)})$) are respectively gathered in matrices $\mathbf{\Pi}$ and $\mathbf{\Pi}^{(-j)}$.

\begin{figure}[t!]
	\begin{center}
		\includegraphics[scale = 0.5]{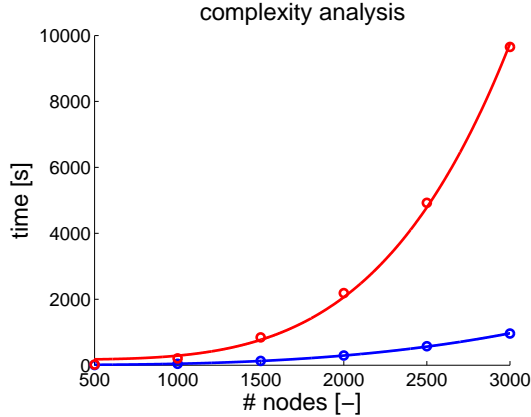}		
		\caption{Empirical complexity analysis. The overall complexity for BPC is $O(n^4)$ (a matrix inversion per node) and for BPCf $O(n^3)$ (only one matrix inversion plus fast updates). Notice that for a 3000-nodes graph, the saving factor is larger than 10. The CPU is a simple Intel(R) Core(TM) i5-4310 at 2.00 GHz with 8 Go RAM and the programming language is Matlab. No sparse, approximate, or optimized implementation were used so that a matrix inversion typically takes a second for a network of 3000 nodes.}
		\label{Complex}
		\end{center}
		\end{figure}

An elementary study of the empirical time complexity of the two versions BPC and BPCf is reported in Figure \ref{Complex}. Recall that the overall complexity for BPC is $O(n^4)$ and $O(n^3)$ for BPCf. For a 3000-nodes graph, the saving factor is greater than 10. Notice that no sparse, approximate, or optimized, implementation were used in the study. The CPU is a simple Intel(R) Core(TM) i5-4310 at 2.00 GHz with 8 Go RAM and the programming language is Matlab.


\begin{algorithm}[t!]
\caption{\small{Computing the bag of hitting paths criticality of $G$.}}

\algsetup{indent=2em, linenodelimiter=.}

\begin{algorithmic}[1]
\small
\REQUIRE $\,$ \\
 -- A graph $G$ containing $n$ nodes. \\
 -- The $n\times n$ adjacency matrix $\mathbf{A}$ associated to $G$, containing affinities.\\
 -- The $n\times n$ cost matrix $\mathbf{C}$ associated to $G$ (usually, the costs are the inverse of the affinities, but other choices are possible).\\
 -- The inverse temperature parameter $\theta$.\\
 
\ENSURE $\,$ \\
 -- The $n \times 1$ bag-of-paths criticality vector $\mathbf{cr}$ containing the change in the probability distribution
of picking a path starting in node $i$ and ending in node $j$ when a node is deleted.\\

\STATE $\mathbf{D} \leftarrow \mathbf{Diag}(\mathbf{A}\mathbf{e})$ \COMMENT{the row-normalization matrix; $\mathbf{e}$ is a column vector full of 1s} \\
\STATE $\mathbf{P}^{\mathrm{ref}} \leftarrow \mathbf{D}^{-1} \mathbf{A}$ \COMMENT{the reference transition probabilities matrix} \\
\STATE $\mathbf{W} \leftarrow \mathbf{P}^{\mathrm{ref}}\circ\exp\left[-\theta\mathbf{C}\right]$ \COMMENT{elementwise exponential and multiplication $\circ$} \\
\STATE $\mathbf{Z} \leftarrow (\mathbf{I}-\mathbf{W})^{-1}$ \COMMENT{the fundamental matrix} \\
\FOR[compute criticality for each node $j$]{$j=1$ to $n$}
\STATE $\mathbf{z}_{j}^{\mathrm{r}} \leftarrow \mathbf{e}_{j}^{\text{T}} \mathbf{Z}$ and
$\mathbf{z}_{j}^{\mathrm{c}} \leftarrow \mathbf{Z} \mathbf{e}_{j}$  \COMMENT{copy row $j$ and column $j$ of $\mathbf{Z}$}


\STATE $\mathbf{Z}' \leftarrow \mathbf{Z} - \mathbf{e}_{j} \mathbf{z}_{j}^{\mathrm{r}} - \mathbf{z}_{j}^{\mathrm{c}} \mathbf{e}_{j}^{\text{T}} + z_{jj}\mathbf{e}_{j}\mathbf{e}_{j}^{\text{T}}$ \COMMENT{set row $j$ and column $j$ of $\mathbf{Z}$ to $0$ for disregarding paths starting and ending in $j$, but still passing through $j$}   
\STATE $\mathbf{\Pi} \leftarrow \dfrac{\mathbf{Z}'}{\mathbf{e}^{\text{T}} \mathbf{Z}' \mathbf{e}}$ \COMMENT{normalize in order to obtain the bag-of-paths probability matrix whose support is now $\mathcal{V} \setminus j$}
\STATE $\mathbf{Z}^{(-j)} \leftarrow \mathbf{Z} - \dfrac{\mathbf{z}_{j}^{\mathrm{c}} \mathbf{z}_{j}^{\mathrm{r}}}{z_{jj}}$ \COMMENT{update of matrix $\mathbf{Z}$ when removing row $j$ from $\mathbf{W}$}
\STATE $\mathbf{\Pi}^{(-j)} \leftarrow \dfrac{\mathbf{Z}^{(-j)}}{\mathbf{e}^{\text{T}} \mathbf{Z}^{(-j)} \mathbf{e}}$
\COMMENT{normalize in order to obtain the corresponding bag-of-paths probabilities after deletion of row $j$ of $\mathbf{W}$}
\STATE Remove both row $j$ and column $j$ from $\mathbf{\Pi}$ and $\mathbf{\Pi}^{(-j)}$
\STATE $\boldsymbol{\pi} \leftarrow \vec{\mathbf{\Pi}}$ and $\boldsymbol{\pi}^{(-j)} \leftarrow \vec{\mathbf{\Pi}^{(-j)}}$
\COMMENT{stack probabilities into column vectors by using the vec operator}
\STATE $\mathrm{cr}_{j} \leftarrow (\boldsymbol{\pi}^{(-j)})^{\text{T}} \log(\boldsymbol{\pi}^{(-j)} \div \boldsymbol{\pi})$ \COMMENT{compute Kullback-Leibler divergence with $\div$ being the elementwise division. It is assumed that $0 \, \log 0 = 0$ and $0 \, \log(0/0) = 0$}
\ENDFOR
\RETURN \textbf{cr}
\end{algorithmic} \label{Alg_bag_of_paths_criticality01} 
\end{algorithm}


\subsection{Illustrative example}
\label{Ill}

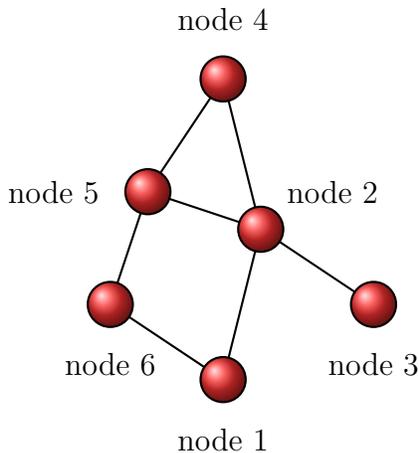
\begin{figure}
\centering
\begin{tikzpicture}[thick]

%
				
		\draw[<->] (1.5, 0) -- (2, 2);
		\draw[<->] (2, 2) -- (3.5, 1);
		\draw[<->] (2, 2) -- (1.5, 4);
		\draw[<->] (1.5, 4) -- (0.5, 2.5);
		\draw[<->] (0.5, 2.5) -- (2, 2);
		\draw[<->] (0.5, 2.5) -- (0, 1);
		\draw[<->] (1.5, 0) -- (0, 1);

    \filldraw[ball color=red!80,shading=ball] (1.5,0) circle
        (0.3cm) node[below=0.5cm]{node 1};
		\filldraw[ball color=red!80,shading=ball] (2,2) circle
        (0.3cm) node[above=0.5cm,right=0.2cm]{node 2};
		\filldraw[ball color=red!80,shading=ball] (3.5,1) circle
        (0.3cm) node[below=0.5cm]{node 3};
		\filldraw[ball color=red!80,shading=ball] (1.5,4) circle
        (0.3cm) node[above=0.5cm]{node 4};
		\filldraw[ball color=red!80,shading=ball] (0.5,2.5) circle
        (0.3cm) node[left=0.5cm]{node 5};
		\filldraw[ball color=red!80,shading=ball] (0,1) circle
        (0.3cm) node[below=0.5cm]{node 6};

\end{tikzpicture} 
\caption{A small toy graph. The (rounded) BPC value for each node is 6.3, 8.5, 5.5, 6.2, 7.1, 6.3, respectively. It corresponds to the node ranking 2, 5, 6, 1, 4, 3, which seems legit. Conversely, WIE succeeds to identify node 2 as the most critical, but the second node in the ranking is node 3, which looks counter-intuitive.}
\label{IllGraph}
\end{figure}

A small toy graph, depicted on Figure \ref{IllGraph}, is now used as an illustrative example. This graph has six nodes: the (rounded) BPC value for each node is 6.3, 8.5, 5.5, 6.2, 7.1, 6.3, respectively. It corresponds to the node ranking 2, 5, 6, 1, 4, 3 (where the largest score defines the most critical node), which seems legit. The WIE criticality succeeds to identify node 2 as the most critical, but the second node in the ranking is node 3, which looks counter-intuitive. 

\section{Experimental comparisons}
\label{Exp}

\begin{table*}[ht]
\caption{List of all measures compared in this study, together with their type, acronym and parameter. If a measure depends on a parameter, tested values as well as the most frequent value (mode) are reported. Notice that Shortest Path and Random Walk Betweenness algorithms are fast, optimized, versions. The other algorithms were implemented in Matlab, as described in Section \ref{RelW}. Further notice that the Matlab implementation of the matrix exponential is very efficient (it is used in Estrada's node betweenness).} 
\begin{center}
\begin{tiny}
\makebox[\textwidth]{
\begin{tabular}{l|c|c|c|c|c|c|c}

\textbf{Name} &\textbf{Type} &  \textbf{Acronym} & \textbf{Description} & \textbf{Param.} & \textbf{Tested values} & \textbf{Mode} & \textbf{Time}\\    
\hline
Baseline		 								& -						& BL 			& Subsection \ref{Deco} 			& none 			& - 										& - 						& $<10^{-3}s$ \\
Edge Connectivity 					& Node Betw. 	& EC 			& See Eq. \ref{EC} 						& none 			& - 										& - 						& $<10^{-3}s$ \\
Shortest Path Betweenness	 	& Node Betw. 	& SPB 		& See Eq. \ref{SPB} 					& none 			& - 										& - 						& $0.8s$ \\
Random Walk Betweenness  		& Node Betw. 	& RWB 		& Subsection \ref{Bet} 				& none 			& - 										& - 						& $1s$ \\
Estrada Index						 					& Node Betw. 	& EST 		& See Eq. \ref{EST} 					& none 			& - 										& - 						& $0.6s$ \\
Wehmuth's K				 					& Node Crit. 	& WK			& See Eq. \ref{WK} 						& $h$    		& [1,2,3,4,5,6] 				& 1 (28\%) 			& $342s$ \\
Klein Index							 					& Node Crit. 	& KLE 		& See Eq. \ref{KLE} 					& none 			& - 										& - 						& $1634s$ \\
Wiener Index			 					& Graph Crit. & WIE 		& See Eq. \ref{WIE} 					& none 			& - 										& - 						& $375s$ \\
Kirchhoff Index							& Graph Crit.	& KIR 		& See Eq. \ref{KIR} 					& none 			& - 										& - 						& $884s$ \\
Kemeny Index								& Graph Crit.	& KEM 		& See Eq. \ref{KEM} 					& none 			& - 										& - 						& $1000s$ \\
Shield Value		 						& Graph Crit.	& SHV 		& See Eq. \ref{SHV} 					& none 			& - 										& - 						& $182s$ \\
Bag-of-Paths criticality (fast version)	& Node Crit.	& BPCf 	& See Eq. \ref{Eq_bag_hitting_paths_criticality01} & $\theta$ & $10^{[-6,-3,-2,-1,0,1]}$ & 10 (44\%) 	& $42s$ \\
Bag-of-Paths criticality (standard version)	& Node Crit.	& BPC 	& See Eq. \ref{Eq_Sherman_criticality01}& $\theta$ & $10^{[-6,-3,-2,-1,0,1]}$ & 1 (39\%) 		& $205s$ \\

\end{tabular}
}
\end{tiny}
\label{Accro}
\end{center} 
\end{table*}

In this section, the bag-of-paths criticalities (both the exact one (BPC) and the fast approximate one (BPCf)) and the other centrality measures introduced in Section \ref{RelW} are computed (see Table \ref{Accro} for a reminder) on the two types of graphs described in subsection \ref{Data}. To do so, we followed a common methodology \cite{albert-2000error,holme-2002,Estrada-2006,Petreska-2010,Santiago-2009} described in subsection \ref{Deco} and we report first a simple correlation analysis between rankings in subsection \ref{Corr}. Then, results are compared and discussed in subsection \ref{ResDis}.

\subsection{Datasets}
\label{Data}
We used two well-known graph generators \cite{Barabasi-99,Bollobas-2001} to build a set of 200 graphs: 100 are generated using Erd\H os-R\'{e}nyi's model and an additional 100 using Albert-Barab\'asi's model. Each of these models has different variants; the one we used is described below. The number of nodes is set randomly for each graph between 5 and 500.
\begin{itemize}
\item \textbf{Erd\H os-R\'{e}nyi (ER) Graph Generator \cite{Bollobas-2001}.}
This model is also called the Poisson random graph generator because it generates a random graph with a Poisson node degree distribution. This type of graph is often used to study theoretical properties and behavior of networks \cite{Newman-2010}. A parameter $p\in {]0,1]}$ is required. The model first generates an upper triangular random matrix (zeros on diagonal), then, for each entry of the matrix, it puts a $0$ if the entry is smaller than $p$, and $1$ otherwise. Then the matrix is symmetrized using $\mathbf{A}+\mathbf{A}^\text{T}$. For our experiments, $p$ was set to a random value for each graph, with $p \in {]0,1/2]}$. \\

\item \textbf{Albert-Barab\'asi (AB) Graph Generator \cite{Barabasi-99}.}
The model generates a random graph with a power law degree distribution. This kind of network is often observed in natural and human-generated systems, including the world wide web, citation networks, and social networks \cite{Newman-2010}. An integer parameter $m$ is required. The model begins with an initial connected network of $m+1$ nodes. Then, new nodes are added to the network, one at a time. Each new node is connected to $m$ existing nodes with a probability that is proportional to the current degree of each node. The procedure stops when the desired number of nodes is reached. Heavily linked nodes (``hubs") tend to quickly accumulate even more links: the new nodes have a ``preference" to attach themselves to these already heavily linked nodes. For our experiments, $p$ was set to a random value for each graph with $m \in \{1,2,3,4,5,6\}$. Many ``natural" networks in real life behave like AB graphs (see for example \cite{albert-2000error} and citations inside).
\end{itemize}

\subsection{Disconnection strategies}
\label{Deco}

\begin{table*}[ht]
\caption{Results obtained with the disconnection strategies described in Subsection \ref{Deco}. The \emph{Friedman/Nemenyi ranking} over 100 graphs (AB and ER), according to two disconnection strategies (single ranking and 100-ranking), is presented, together with the mean $\pm$ standard deviation of the obtained RBCC \emph{area under the curve} (AUC). Concerning the ranking, the Critical Difference is equal to 1.82, meaning that a measure is significantly better than another if their rank difference is larger than this amount. For the ranking, the larger is the better whereas, for AUC, smaller is better. In each column, the methods in bold are the best ones or are not significantly different from the overall best one.}
\begin{center}
\begin{tiny}
\makebox[\textwidth]{
\begin{tabular}{lll|lll|lll|lll}

		\multicolumn{3}{c|}{100 AB graphs: single ranking} & \multicolumn{3}{c|}{100 ER graphs: single ranking}& \multicolumn{3}{c|}{100 AB graphs: 100-ranking} & \multicolumn{3}{c}{100 ER graphs: 100-ranking}  \\
		\hline
		measure &ranking & AUC & measure &ranking & AUC & measure &ranking & AUC & measure &ranking & AUC  \\
		\hline
			\textbf{BPC} &\textbf{11.750}  &\textbf{0.3092 $\pm$0.163} & 
			\textbf{BPC} &\textbf{12.355} &\textbf{0.8634 $\pm$0.180} & 
			\textbf{BPCf}  &\textbf{11.640}  &\textbf{0.3174 $\pm$0.155} & 
			\textbf{BPC} &\textbf{12.555} &\textbf{0.7936 $\pm$0.161} \\ 
			
			\textbf{BPCf}  &\textbf{11.285} &\textbf{0.3103 $\pm$0.164} & 
			       {BPCf}  &       {10.250}  &       {0.8773 $\pm$0.182} & 
			\textbf{BPC} &\textbf{11.370}  &\textbf{0.3185 $\pm$0.156} & 
			\textbf{BPCf}  &\textbf{11.270}  &\textbf{0.8063 $\pm$0.163} \\ 
			
			\textbf{RWB}  &\textbf{10.425} &\textbf{0.3158 $\pm$0.167} & 
			       {SPB}  &       {9.590}   &       {0.8851 $\pm$0.167} & 
			\textbf{WK}   &\textbf{10.375} &\textbf{0.3249 $\pm$0.159} & 
			       {RWB}  &       {9.095}  &       {0.8186 $\pm$0.160} \\
			
			       {KIR}  &9.435           &0.4550 $\pm$0.255 & 
			       {RWB}  &9.365           &0.8827 $\pm$0.175 & 
			       {EC}   &8.645           &0.3392 $\pm$0.162 & 
			       {KIR}  &8.630           &0.8405 $\pm$0.138 \\ 
			
			       {WK}   &8.805           &0.3246 $\pm$0.175 & 
			       {KIR}  &8.575           &0.8954 $\pm$0.150 & 
			{RWB}  &{8.475}  &{0.3427 $\pm$0.162} & 
			       {WK}   &8.275           &0.8215 $\pm$0.163 \\ 
			
			       {SPB}  &8.205           &0.3283 $\pm$0.172 & 
			       {WK}   &7.550           &0.8937 $\pm$0.168 & 
			{EST}  &{8.090}   &{0.3423 $\pm$0.167} & 
			       {SPB}  &7.955           &0.8258 $\pm$0.156 \\ 
						
			       {EC}   &7.815           &0.3276 $\pm$0.176 & 
			       {EC}   &7.290           &0.8959 $\pm$0.165 & 
			{SPB}  &{7.510}   &{0.3523 $\pm$0.164} & 
			       {EC}   &7.730            &0.8272 $\pm$0.159 \\
			
			{KLE}  &{6.940}   &{0.3577 $\pm$0.208} & 
			{WIE}  &{6.610}   &{0.9112 $\pm$0.131} & 
			{KLE}  &{5.945}  &{0.3740 $\pm$0.182} & 
			{KEM}  &{6.280}   &{0.8467 $\pm$0.143} \\
			
			{WIE}  &{4.385}  &{0.5188 $\pm$0.242} & 
			{KEM}  &{5.665}  &{0.9092 $\pm$0.145} & 
			{KIR}  &{5.290}   &{0.5232 $\pm$0.238} & 
			{WIE}  &{5.530}   &{0.8719 $\pm$0.112} \\
			
			{KEM}  &{4.260}   &{0.5226 $\pm$0.246} & 
			{EST}  &{4.230}   &{0.9113 $\pm$0.156} & 
			{KEM}  &{5.230}   &{0.5172 $\pm$0.228} & 
			{EST}  &{5.410}   &{0.8396 $\pm$0.156} \\
			
			{EST}  &{3.620}   &{0.4666 $\pm$0.224} & 
			{SHV}  &{3.780}   &{0.9207 $\pm$0.129} & 
			{SHV}  &{4.005}  &{0.4179 $\pm$0.169} & 
			{SHV}  &{3.730}   &{0.8640 $\pm$0.134} \\ 
			
			{SHV}  &{2.585}  &{0.5035 $\pm$0.185} & 
			{BL}   &{3.015}  &{0.9366 $\pm$0.104} & 
			{WIE}  &{2.635}  &{0.5995 $\pm$0.232} & 
			{KLE}  &{2.790}   &{0.8771 $\pm$0.150} \\
			
			{BL}   &{1.490}  &{0.7078 $\pm$0.193} & 
			{KLE}  &{2.725}  &{0.9273 $\pm$0.134} & 
			{BL}   &{1.795}  &{0.6380 $\pm$0.194} & 
			{BL}   &{1.910}   &{0.8986 $\pm$0.120} \\

\label{TABdeco}
\end{tabular}
}
\end{tiny}
\end{center}
\end{table*}

\begin{table*}[t]
\caption{Another perspective on the results obtained with the disconnection strategies described in Subsection \ref{Deco}. The \emph{Friedman/Nemenyi ranking} over 100 graphs (AB or ER) is presented, together with the mean $\pm$ standard deviation of the RBBC \emph{area under the curve} (AUC). Here, both strategies (single ranking and 100-ranking) are analyzed together. Concerning the ranking, the Critical Difference is equal to 3.97, meaning that a measure is significantly better than another if their rank difference is larger than this amount. For the ranking, the larger is the better while for AUC, the smaller is the better.  In each column, the methods in bold are the best ones or are not significantly different from the overall best one.}
\begin{center}
\begin{tiny}
\makebox[\textwidth]{
\begin{tabular}{lll|lll}

		\multicolumn{3}{c|}{100 AB graphs} & \multicolumn{3}{c}{100 ER graphs}\\
		\hline
		measure &ranking & AUC & measure &ranking & AUC  \\
		\hline
		\textbf{1-BPC}  	&\textbf{23.185} 	&0.30923 $\pm$0.16319 	& \textbf{100-BPC} &\textbf{25.190} 	&0.79360 $\pm$0.16064 \\
    \textbf{1-BPCf}  	&\textbf{22.575} 	&0.31030 $\pm$0.16351 	&\textbf{100-BPCf}  	&\textbf{23.955}	&0.80633 $\pm$0.16343 \\
    \textbf{1-RWB}  	&\textbf{21.045} 	&0.31579 $\pm$0.16704 	&\textbf{100-RWB}  	&\textbf{21.660} 	&0.81857 $\pm$0.15963 \\
    \textbf{100-BPCf}  &\textbf{20.355} &0.31740 $\pm$0.15509 		&\textbf{100-WK}  	&20.880 					&0.82151 $\pm$0.16311 \\
    \textbf{100-BPC}	&\textbf{20.075} 	&0.31847 $\pm$0.15584 	&\textbf{100-KIR}  	&20.440 					&0.84045 $\pm$0.13762 \\
    \textbf{1-WK}  		&18.875 					&0.32461 $\pm$0.17465 	&\textbf{100-SPB}  	&20.250						&0.82587 $\pm$0.15663 \\
    \textbf{1-KIR}  	&18.565 					&0.45500 $\pm$0.25486 	&\textbf{100-EC}  	&20.200 					&0.82723 $\pm$0.15883 \\
    \textbf{100-WK}  	&18.235 					&0.32488 $\pm$0.15874 	&\textbf{100-KEM}  	&18.275 					&0.84668 $\pm$0.14274 \\
    \textbf{1-SPB}  	&17.665 					&0.32831 $\pm$0.17209 	&\textbf{100-EST}  	&17.790						&0.83957 $\pm$0.15614 \\
    \textbf{1-EC}  		&17.585 					&0.32764 $\pm$0.17627 	&\textbf{100-WIE}  	&16.420						&0.87185 $\pm$0.11230 \\
    \textbf{100-EC}  	&15.525 					&0.33924 $\pm$0.16238 	&\textbf{100-SHV}  	&15.375 					&0.86398 $\pm$0.13363 \\
    \textbf{100-RWB}  &15.120						&0.34273 $\pm$0.16239 	&\textbf{1-BPC}  	&14.855 					&0.86344 $\pm$0.18018 \\
    \textbf{100-EST}  &14.890						&0.34227 $\pm$0.16667 	&\textbf{100-KLE}  	&13.595 					&0.87705 $\pm$0.15034 \\
    \textbf{1-KLE}  	&14.780						&0.35771 $\pm$0.20768 	&\textbf{1-BPCf}  		&12.160						&0.87733 $\pm$0.18177 \\
    \textbf{100-SPB}  &13.395 					&0.35227 $\pm$0.16436 	&\textbf{100-BL}  	&11.245 					&0.89857 $\pm$0.11965 \\
    \textbf{100-KLE}  &10.675 					&0.37401 $\pm$0.18217 	&\textbf{1-SPB}  		&11.120						&0.88511 $\pm$0.16666 \\
    \textbf{100-KIR}  &10.085 					&0.52323 $\pm$0.23805 	&\textbf{1-RWB}  		&11.005 					&0.88265 $\pm$0.17446 \\
    \textbf{100-KEM}  &10.065 					&0.51721 $\pm$0.22753 	&\textbf{1-KIR}  		&9.655						&0.89537 $\pm$0.15000 \\
    \textbf{1-WIE}  	&8.685 						&0.51884 $\pm$0.24191 	&\textbf{1-WK}  		&8.900						&0.89371 $\pm$0.16805 \\
    \textbf{1-KEM}  	&8.470						&0.52258 $\pm$0.24621 	&\textbf{1-EC}  		&8.470 					  &0.89593 $\pm$0.16539 \\
    \textbf{100-SHV}  &7.905 						&0.41789 $\pm$0.16913 	&\textbf{1-WIE}  		&7.290 					  &0.91115 $\pm$0.13097 \\
    \textbf{1-EST}  	&7.300						&0.46664 $\pm$0.22439 	&\textbf{1-KEM}  		&6.405 					  &0.90924 $\pm$0.14499 \\
    \textbf{100-WIE}  &5.600						&0.59947 $\pm$0.23205 	&\textbf{1-EST}  		&5.040 					  &0.91128 $\pm$0.15548 \\
    \textbf{1-SHV}  	&4.815 						&0.50353 $\pm$0.18480		&\textbf{1-SHV}  		&4.390 					  &0.92066 $\pm$0.12886 \\
    \textbf{100-BL}  	&3.190						&0.63796 $\pm$0.19396 	&\textbf{1-BL}  		&3.300						&0.93658 $\pm$0.10436 \\
    \textbf{1-BL}  		&2.340						&0.70782 $\pm$0.19250		&\textbf{1-KLE}  		&3.135						&0.92732 $\pm$0.13371 \\
		
\label{TABdecoll}
\end{tabular}
}
\end{tiny}
\end{center}
\end{table*}

\begin{table*}[ht]
\caption{Selected parameters for the disconnection strategies described in Subsection \ref{Deco}. Note that only WK, BPCf and BPC need a parameter to be tuned. Bold values show the maximum per task and per measure.}
\begin{center}
\begin{tiny}
\makebox[\textwidth]{
\begin{tabular}{c|l|c|c|c|c|c}


	  	& & 100 AB graphs: & 100 ER graphs: & 100 AB graphs: & 100 ER graphs: & sum over\\
	  	measure & parameter value & single ranking & single ranking & 100-ranking & 100-ranking & the 4 tasks\\
	  \hline
	  \multirow{6}{0.5cm}{WK}
			 &\quad $h=1$ & \textbf{28} & 15 & \textbf{52} & 14 & \textbf{111} \\
			 &\quad  $h=2$ & 7  & \textbf{68} & 6  & 7  & 87 \\
			 &\quad  $h=3$ & 27 & 15 & 8  & 23 & 70 \\
			 &\quad  $h=4$ & 27 & 0  & 9  & \textbf{26} & 60 \\
			 &\quad  $h=5$ & 8  & 2  & 13 & 16 & 40 \\
			 &\quad  $h=6$ & 3  & 0  & 12 & 14 & 32 \\
	  \hline
	  	\multirow{6}{0.5cm}{BPCf}
	  	&\quad  $\theta = 10^{-6}$ & 6  & 24 & \textbf{29} & 10 & 61 \\
	  	&\quad  $\theta = 0.001  $ & 6  & 0  & 9 & 3  & 18 \\
	  	&\quad  $\theta = 0.01   $ & 6  & 1  & 12 & 5  & 24 \\
	  	&\quad  $\theta = 0.1    $ & 8  & 2  & 19 & 14 & 44 \\
	  	&\quad  $\theta = 1      $ & 18 & 6  & 22 & 29 & 76 \\
	  	&\quad  $\theta = 10     $ & \textbf{56} & \textbf{67} & 9 & \textbf{39} & \textbf{177} \\
	  \hline
	  	\multirow{6}{0.5cm}{BPC}
	  	&\quad  $\theta = 10^{-6}$ & 16 & 21 & 8 & 15 & 60 \\
	  	&\quad  $\theta = 0.001  $ & 6  & 1  & 4  & 4  & 14 \\
	  	&\quad  $\theta = 0.01   $ & 17 & 3  & 2  & 1  & 23 \\
	  	&\quad  $\theta = 0.1    $ & 24 & 8  & 17 & 18 & 66 \\
	  	&\quad  $\theta = 1      $ & 12 & \textbf{52} & \textbf{49} & \textbf{45} & \textbf{156} \\
	  	&\quad  $\theta = 10     $ & \textbf{25} & 15 & 20 & 17 & 81
\label{TABparam}
\end{tabular}
}
\end{tiny}
\end{center}
\end{table*}

To study the performances of the different centrality/criticality measures, we simulate the effect of network attacks consisting in deleting its nodes sequentially in the order provided by the measure -- the most critical nodes being deleted first. This is a natural way of assessing node criticality \cite{albert-2000error,holme-2002}. We then record, for each network and each measure, the results of this sequential node deletion by measuring its gradual impact on network connectivity. A good criticality measure hurts most the network by, e.g., disconnecting it in several connected components, each preferably having an equal size -- a balanced partition.

In practice, we first compute a criticality ranking of all nodes according to each different centrality/criticality measure introduced in the previous section. This ranking can be achieved in two different way: (1) it is computed once for all from the whole graph $G$ (one \textit{single ranking}), or (2) it is re-computed after each node deletion. With this last option, the centrality/criticality measures must be re-computed $n-1$ times which is time-consuming. We therefore decided to update the ranking only 100 times in total (except, obviously, for graphs with $n<100$ nodes). This last option will be referred to as \textit{100-rankings}.

Recall that, to evaluate the criticality of a node $j$ with respect to a global graph criticality measure, the difference between the graph criticality of $G \setminus j$ and the global graph $G$ criticality is computed (see Equation (\ref{Eq_global_graph_criticality})). 

\begin{figure}[t!]
\begin{center}
		\includegraphics[scale = 0.50]{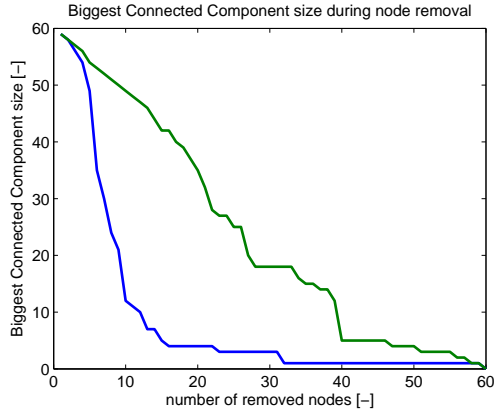}
		\caption{Example of Biggest Connected Component size recorded when nodes are removed following criticality rankings. The network is an Albert-Barab\'asi (AB) 60-nodes graph. The two criticality rankings are BPC (lower curve) and BL (upper curve) and are computed once before starting to remove nodes. The BPC ranking is more efficient in detecting the critical nodes, as their removal quickly disconnects the network.}
		\label{BCCvsNR}
\end{center}
\end{figure}

Once those node rankings have been computed for each measure, the simulated attack can start. Nodes are deleted in decreasing order of criticality. After each node deletion, the \emph{Biggest Connected Component} size (BCC), i.e., the number of nodes contained in the largest connected component, is recorded \cite{albert-2000error,holme-2002}. The smaller this value, the more effective the attack and thus the more effective the criticality index (see Figure \ref{BCCvsNR} for an example). This performance measure quantifies to which extend the network is decomposed in several balanced parts (no ``giant" component is left). If, for example, the node deletion strategy (the criticality ranking) is very inefficient, and it never disconnects the network, the BCC only decreases by one unit at a time. On the contrary, if it cuts the network into two equally sized parts, the BCC is divided by two, which corresponds to a large decrease.

By further normalizing with respect to the size of the graph, that is, dividing BCC by the current number of nodes, we get the \emph{Relative Biggest Connected Component} size (RBCC) which will be the performance indicator used in the experiments. It is then possible to draw a plot of RBCC versus the number of deleted nodes $(1,2,3,\dots,n)$ \cite{albert-2000error,holme-2002}. Then, to summarize those plots, we sum up the \emph{Area Under the Curve} (AUC). The smaller this AUC, the better the method since the deletion of the most critical nodes (according to the ranking) quickly disconnects the network into balanced components, leading to smaller RBCC (see the illustrative example in Figure \ref{BCCvsNR}).

Finally, we report our results as follows: we perform a Friedman/Nemenyi test \cite{Demsar2006} and, in addition, we also compute the mean and the standard deviation of the AUC across all of the AB and ER generated graphs, providing more detailed results. Results can be found on Table \ref{TABdeco}; the higher the ranking, the better the measure. 

If a parameter is present, it is tuned as follows: for each graph, a range of values is tested and the best one is chosen for the disconnection experiment (the size of the graph can influence the parameter choice). This reflects the case of a real attack (we assume that the attacker has access to the network structure and can test the effect of different parameters). Parameters could be tuned again after each node deletion, but it would be too computationally intensive, so we did not investigate this approach. For information, best values of parameters $h$ and $\theta$ are reported on Table \ref{TABparam}.

For comparison, we also consider the case where nodes are simply removed at random and independently (BL for baseline). It corresponds to a random ``failure" random or ``attack", which has been studied theoretically in the literature (see \cite{albert-2000error} for an example).

\subsection{Preliminary exploration: correlation analysis}
\label{Corr}

\begin{figure}[t!]
\begin{center}
		\includegraphics[scale = 0.60]{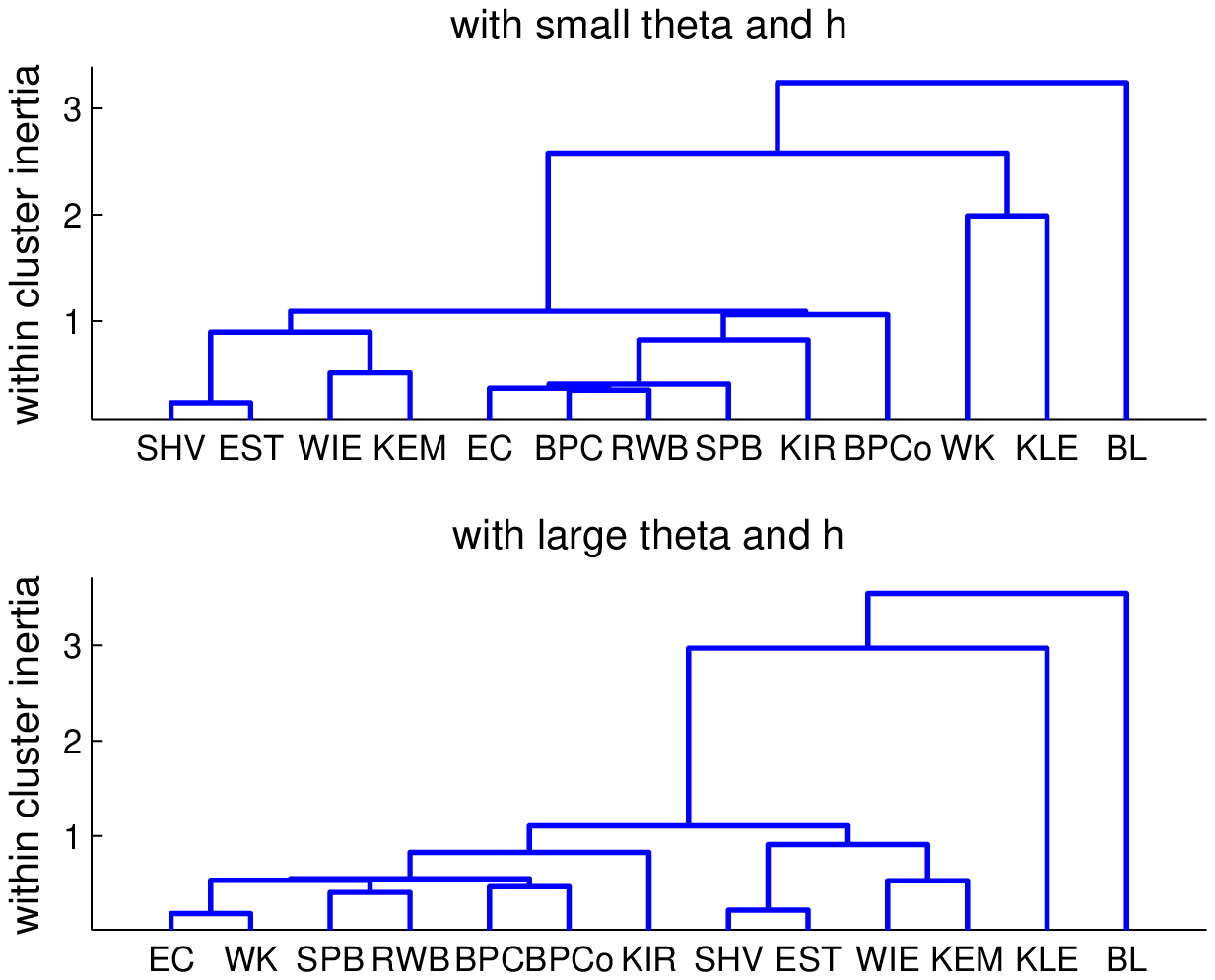}
		\caption{Ward dendrograms of studied criticality measures. Distances are based on Kendall's correlation of Table \ref{TABcorr}. The smaller the height (Y-axis) of joining branches, the closer the measures. As BPCf, BPC and WK depend on a parameter, two cases are considered: a larger value of the parameters and a smaller value. The small $\theta$ and $h$ are $10^{-6}$ and 1, respectively, while the larger $\theta$ and $h$ are 10 and 6.}
		\label{dendro}
\end{center}
\end{figure}

 The different centrality/criticality measures were first compared by computing two Kendall's correlation tests between each ranking. This is reported on Table \ref{TABcorr} for both a small and a larger value of the parameters of our centrality/criticality measures: $\theta$ (BPCf and BPC) and $h$ (WK). The small $\theta$ and $h$ were set to $10^{-6}$ and 1, respectively, while the larger $\theta$ and $h$ were $10$ and 6. To summarize and to make things more visual, dendrograms were built above with a Ward hierarchical clustering ADD REF based on Kendall's correlation matrices (Figure \ref{dendro}).

\begin{table*}[t]
\caption{Mean Kendall's correlation between selected measures over our 200 graphs. Above the main diagonal: with larger $\theta$ and $h$. Below the main diagonal: small $\theta$ and $h$.}
\begin{center}
\begin{tiny}

\makebox[\textwidth]{
\begin{tabular}{l|ccccccccccccc}

	& EC  &  SPB  &  SHV  &  WK  &  BPCf  &  WIE  &  KIR  &  KLE  &  EST  &  BL  &  KEM  &  RWB  &  BPC\\
		\hline
    
  EC  &  \textbf{1.0000}  &  0.8784  &  0.7352  &  0.9745  &  0.8525  &  0.7088  &  0.7562  &  0.3452  &  0.7761  & -0.0030  &  0.6848  &  0.8895  &  0.7637 \\
  SPB &  0.8784  &  \textbf{1.0000}  &  0.6367  &  0.8555  &  0.8169  &  0.6666  &  0.7365  &  0.3987  &  0.6732  & -0.0047  &  0.6420  &  0.8790  &  0.7652 \\
  SHV &  0.7352  &  0.6367  &  \textbf{1.0000}  &  0.7523  &  0.5391  &  0.7678  &  0.5992  &  0.1036  &  0.9476  & -0.0072  &  0.7074  &  0.6008  &  0.4898 \\
  WK  &  0.3392  &  0.3116  &  0.0958  &  \textbf{1.0000}  &  0.8153  &  0.7470  &  0.7929  &  0.3286  &  0.7850  & -0.0033  &  0.7206  &  0.8766  &  0.7377 \\
  BPCf &  0.9231  &  0.8579  &  0.6607  &  0.2959  &  \textbf{1.0000}  &  0.5761  &  0.7689  &  0.4441  &  0.5600  & -0.0022  &  0.5345  &  0.8695  &  0.8863 \\
  WIE &  0.7088  &  0.6666  &  0.7678  &  0.0744  &  0.6873  &  \textbf{1.0000}  &  0.7821  &  0.1380  &  0.7559  & -0.0073  &  0.8380  &  0.6384  &  0.5694 \\
  KIR &  0.7562  &  0.7365  &  0.5992  &  0.1620  &  0.8185  &  0.7821  &  \textbf{1.0000}  &  0.3303  &  0.5923  & -0.0066  &  0.7345  &  0.8055  &  0.7585 \\
  KLE &  0.3452  &  0.3987  &  0.1036  &  0.3325  &  0.3747  &  0.1380  &  0.3303  &  \textbf{1.0000}  &  0.1314  &  0.0060  &  0.1075  &  0.4055  &  0.4675 \\
  EST &  0.7761  &  0.6732  &  0.9476  &  0.1340  &  0.6864  &  0.7559  &  0.5923  &  0.1314  &  \textbf{1.0000}  & -0.0032  &  0.6956  &  0.6350  &  0.5030 \\
  BL  &  0.0058  &  0.0047  &  0.0037  &  0.0003  &  0.0058  & -0.0016  & -0.0000  &  0.0100  &  0.0033  &  \textbf{1.0000}  & -0.0057  & -0.0050  & -0.0064 \\
  KEM &  0.6848  &  0.6420  &  0.7074  &  0.0847  &  0.6671  &  0.8380  &  0.7345  &  0.1075  &  0.6956  & -0.0005  &  \textbf{1.0000}  &  0.6292  &  0.5038 \\
  RWB &  0.8895  &  0.8790  &  0.6008  &  0.2945  &  0.9093  &  0.6384  &  0.8055  &  0.4055  &  0.6350  &  0.0068  &  0.6292  &  \textbf{1.0000}  &  0.7959 \\
  BPC&  0.7566  &  0.7575  &  0.4538  &  0.3078  &  0.7480  &  0.4820  &  0.6818  &  0.5083  &  0.4753  &  0.0097  &  0.4451  &  0.7872  &  \textbf{1.0000} 
    
\label{TABcorr}
\end{tabular}
}
\end{tiny}
\end{center}
\end{table*}

\subsection{Results and discussion}
\label{ResDis}

First, notice that, when performing a Friedman/Nemenyi test comparing different rankings provided by the methods, the \emph{critical difference} is equal to 1.82, meaning that a measure is considered as significantly better than another if its rank is larger by more than this amount. For three of the four considered tasks (for both disconnection strategies, single ranking and 100-rankings, on Albert-Barab\'asi (AB) graphs and 100-rankings on Erd\H os-R\'{e}nyi (ER) graphs but not for single ranking on ER graphs), the Friedman/Nemenyi test \cite{Demsar2006} cannot conclude that our proposed model (BPC) is better than its approximation, BPCf, and vice versa (Table \ref{TABdecoll}). For the ranking with ER graphs the rankings are significantly different but still close in comparison to other criticalities. It means that the considered approximation seems reasonable, at least on the studied datasets.

We also observe from the same test (Table \ref{TABdecoll}) that BPC is significantly better than all the other tested measures on ER graphs. On AB graphs, it cannot be concluded that BPC is significantly better than RWB in the case where only one ranking is performed (single ranking). This is probably related to the fact that BPC is based on random walks, as RWB. Moreover, if an updated ranking is used instead (100-rankings), then BPC is not significantly better than WK -- while still obtaining better performances. We conclude that the introduced criticality measures (BPC and BPCf) perform well in all contexts as they always perform better (and, most of the time, significantly better) than the competing measures. However, this advantage is not always statistically significant when compared to RWB (single ranking on AB graphs) and WK (100-rankings on AB graphs).

Besides this, we often find the RWB, KIR, WK and SPB measures in the top-5 (Table \ref{TABdecoll}). Notice that the EC (the degree) is quite efficient combined with multiple ranking on AB graphs, given its simplicity. At the bottom of the rankings, KLE, WIE, KEM, EST, and SHV often appear to be even less effective than EC. Since EC is a really obvious measure that can be easily computed, it would certainly be interesting to use EC instead of other, more sophisticated, measures in many settings. EC is quite efficient on AB graphs, if recomputed after each node deletion. It can also be noted that KLE is not performing well on ER graphs (it can even be worse than the random baseline BL, but its mean AUC is still better). We unfortunately do not have a clear explanation of why this is the case.

It is also interesting to identify the most chosen $\theta$ and $h$ parameter from Table \ref{TABparam}. For $h$, it depends on to task to fulfill but the best $h$ value is usually small (1 to 4), and for $\theta$ it is better to take a value between 1 and 10. Notice that BPCf still exhibits the best mean rank when its parameter is fixed (results not presented here; see the discussion at the end of this section).

From Table \ref{TABcorr}, it is clear that WK's correlation with the other measures varies a lot depending of the $h$ value. On the other hand, BPC's and BPCf's correlation with the other measures are less dependent of $\theta$. Notice that it is expected that those measures should be highly correlated with RWB and EC when $\theta$ is small and with SPB when $\theta$ is large, as the bag-of-paths betweenness does \cite{Kivimaki-2016}. However, we observe that the criticality measures BPC and BPCf are still more correlated with RWB when $\theta = 10$.

In Figure \ref{dendro}, we once more notice that the behaviour of WK is strongly dependent of $h$. It turns out that with small $h$, its behavior is similar to KLE. When $h$ is larger, the neighborhood is more and more likely to be close to the whole graph, therefore more and more correlated to EC. As from Table \ref{TABcorr}, BPC's and BPCf's behavior are less sensitive to $\theta$.

From visual inspection of Figure \ref{dendro}, we can identify different clusters of measures:
\begin{itemize}
\item WIE, KEM, SHV and EST form a cluster. This is a bit surprising as these measures are based on different properties of the graph, but still provide relatively similar results. Indeed, WIE is based on shortest paths, KEM is based on random walks, SHV is based on an eigenvalue of $\mathbf{A}$ and EST on paths of different lengths. 
\item SPB, RWB, KIR, EC, BPCf, BPC are part of another cluster. The same observation can be made. If RWB, BPCf and BPC are based on random walks, SPB is based on shortest path and KIR is based on the spectrum of the Laplacian matrix. Notice that SPB, RWB, KIR, BPCf and BPC tend to show good performances on Tables \ref{TABdeco} and \ref{TABdecoll}.
\item KLE looks apart, but is correlated to WK when $h$ is small.
\item Finally, notice that the random baseline BL is the last merged measure in the two cases, which looks natural. 
\end{itemize}

Before closing the discussion, let us comment on the presence of parameters. At first sight, it seems unfair to compare measures depending on a parameter (WH, BPC and BPCf) against measures free of parameter. Recall, however, that the attacker can adapt its behavior to the network structure, so that a parameter monitoring the smoothing scale can be considered as an advantage. Moreover, let us recall two facts about the parameter $\theta$ of BPC and BPCf. First, measures are not very sensitive to the parameter and, second, its optimal value (according to our experiments) is often close to 1 or 10. Therefore, it seems that we could also just fix this parameter. By the way, we reproduced the experiments by setting $\theta = 1$ and it turns out that BPC was still the best measure for three disconnection strategies while the BPCf was the best for the last one (experiments not reported here).

Finally, methods can be sorted (the first been the best one) using Borda score ranking \cite{MPol}:  
\begin{itemize}
	\item If node ranking is updated after each node deletion, independently of the graph type: BPC, BPCf, RWB, WF, EC, SPB/KIR, EST, KEM, KLE, WIE, SHV.
	\item If node ranking is not updated after each node deletion, independently of the graph type: BPC, BPCf, RWB, KIR, SPB, WK, EC, KLE, KEM, WIE EST, SHV.
	\item Finally, independently of the graph type and update factor: BPC, BPCf, RWB, KIR, SPB, WK, EC, KLE, KEM, WIE EST, SHV.
\end{itemize}
These ranking are in concordance with the rest of this Section.

\section{Conclusion}
\label{CCL}

This paper investigated centrality/criticality measures on graphs through a node disconnection analysis and introduced a new criticality measure based on a bag-of-paths framework and its variant: the bag-of-paths criticality and its fast, approximate, version.

Comparisons based on node disconnection simulations performed on a large number of generated graphs show that those two bag-of-paths criticality methods outperform the other considered centrality/criticality measures. Friedman/Nemenyi tests confirm this fact statistically in almost all cases. 

Of course the node disconnection analysis is only a proxy to determine if our criticalities are able to identify ``critical" nodes. Our future work will mainly focus on testing the proposed measures on other tasks and to consider other strategies, such as disconnecting groups of nodes instead of one single node at each time.

Finally, a simple correlation analysis of those measure allowed to identify coherent groups, namely the WIE, KEM, SHV and EST versus the SPB, RWB, KIR, EC, BPCf and BPC (see Table \ref{Accro} for acronyms). It was also shown that the choice of the $\theta$ parameter does not impact much the behavior of our two proposed criticality measures. 

\section{Acknowledgements}

This work was partially supported by the Immediate and the Brufence projects funded by InnovIris (Brussels Region). We thank this institution for giving us the opportunity to conduct both fundamental and applied research.

\section*{References}

\bibliographystyle{elsarticle-num} 
\bibliography{Biblio}





\end{document}